# Pushing the accuracy of on-top functionals with agent-driven supervised learning

*Yuhao Chen*[1], *Dayou Zhang*[2] , *Donald G. Truhlar*[2], *Xiao He*[1,3,4*]

[1]*Shanghai Engineering Research Center of Molecular Therapeutics and New Drug Development, Shanghai Frontiers Science Center of Molecule Intelligent Syntheses, School of Chemistry and Molecular Engineering, East China Normal University, Shanghai, 200062, China;*

[2]*Department of Chemistry, Chemical Theory Center, and Minnesota Supercomputing Institute, University of Minnesota, Minneapolis, MN 55455-0431, USA;*

[3] *Anhui Provincial Key Laboratory of Advanced Catalysis and Energy Materials, Anhui Ultra High Molecular Weight Polyethylene Fiber Engineering Research Center, School of Chemistry and Chemical Engineering, Anqing Normal University, Anqing, 261433, China*

[4]*New York University–East China Normal University Center for Computational Chemistry, New York University Shanghai, Shanghai, 200062, China*

* To whom correspondence should be addressed: xiaohe@phy.ecnu.edu.cn (X.H.)

## Abstract

Multiconfiguration pair-density functional theory (MC-PDFT) combines multiconfigurational reference wave functions with on-top functionals to provide an efficient electronic-structure method for strongly correlated molecular systems. We developed a large language model-based agent for functional development and combined multiple databases into the large, chemically diverse MCDDB26 database. We then used MCDDB26 to develop two on-top functionals, MC26 and COF26. MC26 retains an existing analytical form but substantially improves performance by refitting its linear parameters with optimised weights for the training datasets. Building on this result, we developed COF26, a general hybrid on-top functional with a new analytical form. COF26 provides greater overall accuracy for multireference chemistry than existing on-top functionals. Applications across MCDDB26 show that COF26 accommodates both multiconfigurational and single-configuration densities. It therefore performs well for barrier heights, isomerisation energies, thermochemistry, non-covalent interactions, radical and non-radical chemistry, single-reference and multireference systems, and transition-metal chemistry.

Electronic energies underpin predictions throughout chemistry and materials science, including molecular structures, reaction pathways, spectra and material properties [1]. Density functional theory (DFT) expresses the ground-state energy as a functional of the electron density. It can therefore access broad regions of chemical space that are impractical for high-level wave-function methods [2]. DFT is formally exact [3,4]. In practice, however, Kohn-Sham density functional theory (KS-DFT) remains limited by an unresolved problem. Existing approximations often perform well when one electronic configuration dominates, but can become unreliable when several configurations approach degeneracy [5,6]. Such multiconfigurational situations are widespread in bond dissociation, many transition-metal and actinide compounds [7,8], and electronically excited states [1,9]. In these cases, conventional single-reference density functionals can yield poorly balanced energies and even qualitatively incorrect potential-energy surfaces [10,11]. MC-PDFT combines a multiconfigurational reference wave function with an on-top pair-density functional description of the nonclassical energy. This functional is expressed in terms of the electron density and the on-top pair density [1,12,13]. MC-PDFT thereby retains the multireference character required for strongly correlated systems while recovering dynamic correlation at a cost comparable to KS-DFT. Over the past decade, MC-PDFT has shown considerable promise for bond dissociation energies, reaction barriers, excitation energies and spin-state energetics. It has consequently become a practical framework for chemically complex electronic-structure problems [14–16].

Existing on-top functionals nevertheless leave substantial room for improvement[17]. Many are translated functionals derived from KS-DFT functionals[12,13], which often struggle to describe multireference character accurately[17]. The parametrised MC23 and MC25 functionals [18,19] achieve good accuracy on their training datasets through parameter optimisation. However, they do not retain comparable accuracy across diverse chemical systems and problem classes. These include main-group molecules, transition-metal compounds, single-reference systems, bond dissociation, excited states and spin-state energetics.

Large language model (LLM)-based agents offer a potential means of organising this type of constrained, workflow-level optimisation [20,21]. Previous studies have shown that LLM agents can integrate specialised tools to perform complex tasks in synthesis planning, molecular design and reaction optimisation. They can even support semi-autonomous experimental design and execution on automated laboratory platforms [22–24]. A promising and most recent direction is to use LLM agents to assist functional-dataset development and energy-functional optimisation. Here, we introduce FunctionalAgent for developing MC-PDFT on-top functionals. FunctionalAgent enabled us to construct the Multi-Configuration Diverse Database (MCDDB26). This large and compositionally complex database spans broad chemical space and includes both single-reference and multi-reference systems.

Using FunctionalAgent, we first reoptimised the MC25 functional on MCDDB26 to obtain MC26. Although MC26 retains the same analytical form as MC25, it reduces the training-set MUE and generalises better across benchmark datasets. We introduced a flexible, nonseparable exchange-correlation energy term and developed COF26 with a new analytical form. FunctionalAgent optimised its parameters on MCDDB26 through performance-triggered iterative supervised learning and dataset-weight assignment. Among the on-top functionals examined here, COF26 performs best for strongly correlated systems. It also achieves high accuracy on general benchmarks spanning diverse chemical classes and both single-reference and multireference systems.

# Results

## Agent-driven workflow and functional optimization

To develop the MC26 and COF26 on-top functionals, we constructed FunctionalAgent, an LLM-based agentic system for MC-PDFT (Fig. 1). The system coordinates three main modules. The Active-Space Input Agent analyses multireference character, selects active spaces and generates input files. The QChem Calculation Agent performs and monitors calculations. The Functional Optimization Agent diagnoses performance deficiencies and conducts performance-triggered training-dataset augmentation and functional optimization (Fig. 2; see Methods for technical details). Before full-scale calculations begin, the orchestration agent generates a detailed execution plan for review and approval by human experts. This design combines the processing capacity of an automated pipeline with domain-expert oversight.

The functional form of COF26 and its optimisation by the agent are described in detail in Methods. Here, we provide a brief overview. Following the performance-triggered supervised-learning strategy used to develop CF22D, FunctionalAgent alternates parameter fitting with broader validation on datasets outside the current training set. It diagnoses weaknesses at both the chemical-category and individual-dataset levels. It then adaptively adjusts the training-set composition, dataset weights and regularisation to address the identified problems. When a persistent systematic weakness is detected, challenging datasets or chemically related problem classes are added to the training set. A candidate is selected as the better model only if it improves global accuracy without substantially degrading performance on any key dataset. This criterion prevents gains in aggregate metrics from concealing local degradation in chemically important subsets. These validation datasets guide model development. The closed-loop strategy jointly addresses training accuracy, broader validation performance and robustness across datasets (see Methods for the five-step iterative procedure).

To provide the chemically diverse training and evaluation space required for this optimisation, we assembled MCDDB26 from 150 datasets and 18,954 data points spanning ground-state, excited-state, general-purpose, active-space-robustness, transition-metal and non-covalent-interaction benchmarks (Fig. 3; see Methods and Supplementary Data 1). To characterise FunctionalAgent's optimisation behaviour during the development of MC26 and COF26, we analysed the sampling space generated by performance-triggered supervised learning (Fig. 4). Each sampled functional was evaluated using the overall mean unsigned error (MUE) on the final supervised-learning training set and its mean rank among functionals. Along the MC26 optimisation trajectory, the selected functional lies in a region of low training-set error and low mean rank. MC26 substantially improves training-set performance over MC23 and MC25, which share the same functional form. Relative to MC26, COF26 introduces a nonseparable exchange-correlation term. It also advances markedly through the sampling space compared with MC23 and MC25.

## Benchmark assessment

We evaluated the MC-PDFT functionals on the MC25 training set to assess their quantitative accuracy and consistency across diverse chemical-energy classes containing multireference systems. Two complementary metrics were used to compare the FunctionalAgent-derived functionals with a range of reference methods: rank within each subset and overall MUE across selected datasets. The rank-based analysis assigns equal weight to each subset. It therefore emphasises robustness across chemical classes and reduces biases caused by imbalanced dataset sizes or error scales. The overall MUE instead measures aggregate numerical accuracy. Figure 5 presents a rank heat map for COF26, MC26 and 20 additional DFT- and CASSCF-based reference methods across 26 databases in the overlapping portion of the training sets. The integer ranks are based on MUE, with a lower rank indicating a smaller MUE. Methods are ordered by their mean rank across all datasets. The lower panel shows the overall MUE across the selected ground-state training subsets.

Owing to its nonseparable exchange-correlation term, COF26 ranks above MC26 and MC25 on most datasets. It also attains the best mean rank and lowest overall MUE among all methods. COF26 ranks first on 12 datasets, indicating broad applicability across chemical space containing single-reference and multireference systems. COF26 is substantially more accurate than CASPT2. This result parallels the established single-reference observation that KS-DFT with an optimised modern functional is more accurate than MP2 for most properties. Although it uses the same functional form, MC26 outperforms MC25 and MC23 in both mean rank and overall MUE. This improvement shows that FunctionalAgent-driven supervised learning can effectively adjust dataset weights and regularisation hyperparameters during functional training.

Among recent KS-DFT methods, CF22D is surpassed only by the trained MC-PDFT functionals. The widely used PBE and B3LYP functionals perform relatively poorly on this challenging group of databases. As shown in the lower panel of Fig. 5, tM06-L has a lower overall MUE than M06-L, whereas tτ-HCTH has a higher overall MUE than τ-HCTH. tPBE also outperforms PBE. These results indicate that modern functionals require dedicated reoptimisation within the MC-PDFT framework to achieve high accuracy.

## Excitation energies

COF26 achieves the lowest two-category mean MUE across the examined excited-state training subsets. The comparison covers three same-spin excitation subsets and three spin-changing excitation subsets from QUESTDB. It also includes the TM-SpinSplitting10 and MG-SpinSplitting26 spin-splitting datasets. To compare accuracy across excitation types, we first calculated the arithmetic mean of the subset MUEs within each of two categories. These comprise same-spin excitations and spin-changing excitations plus spin splittings. We then assigned equal weight to the two category means. The resulting MUE for COF26 is 0.1414 eV, slightly below 0.1423 eV for MC25 and 0.1459 eV for MC26. COF26 therefore ranks first among the 11 methods shown.

The advantage of COF26 is concentrated in the QUEST excitation energies. COF26 ranks first for all three spin-changing subsets, with MUEs of 0.08–0.12 eV, whereas MC26 ranks second. COF26 ranks second for all three same-spin subsets, behind only MRPT2, which has MUEs of 0.11–0.14 eV. MC26 ranks third. Both COF26 and MC26 have lower MUEs than MC25 across these six subsets, showing that reparametrisation improved the QUEST excitation energies. The new analytical form and reoptimisation of COF26 further reduce the error for all six subsets. They improve the accuracy of both same-spin and spin-changing excitation energies.

The spin-splitting datasets reveal differences in performance among the methods. For TM-SpinSplitting10, MC25L gives the lowest MUE of 0.09 eV, followed by MC25 at 0.11 eV. The MUEs of MC26 and COF26 are 0.20 and 0.22 eV, respectively. For MG-SpinSplitting26, MRPT2 ranks first with an MUE of 0.14 eV. The corresponding MUEs of MC26 and COF26 are 0.17 and 0.18 eV.

## Chromium dimer: an enduring challenge in quantum chemistry

Despite comprising only two atoms, the chromium dimer $Cr_2$ embodies the complexity of multiple transition-metal bonding and strong electron correlation. The 3*d* and 4*s* shells of the two Cr atoms provide 12 valence electrons, giving a formal sextuple-bond electron count. Its actual bonding, however, combines a short bond length with a relatively weak binding strength. The experimentally derived ground-state potential-energy curve has a narrow well near 1.68 Å and a broad, shallow shoulder near 2.5 Å [25,26]. This unusual profile is associated with the different radial extents of the 3*d* and 4*s* orbitals. The short-range well primarily reflects bonding interactions between the 3*d* orbitals, whereas the stretched-region shoulder is associated with the more diffuse 4*s* orbitals. $Cr_2$ therefore provides a stringent benchmark of whether theoretical methods can describe transition-metal--metal bonding across distinct distance scales [26].

The computational difficulty of $Cr_2$ arises because both static and dynamic correlation are decisive across the entire potential-energy curve. Its singlet ground state dissociates into two high-spin Cr atoms, each containing six unpaired valence electrons. Their spin coupling and near-degenerate configurations give the molecule pronounced multireference character [27]. Electron excitations outside the valence space and core--valence correlation are also important for the binding energy and curve shape. CASSCF treatments of static correlation within the valence space alone cannot reproduce the experimental well near equilibrium. CCSD(T), which is often used to account for dynamic correlation, also struggles to reproduce both a reasonable bond length and the shoulder profile [26]. These dual requirements have driven continued development of multireference perturbation theory, DMRG and selected configuration-interaction methods for $Cr_2$. They also make the choices of active space and basis set critical for quantitative calculations [27].

Figure 7 compares $Cr_2$ bond-dissociation curves based on a CAS(12e,22o) active space. The multiconfigurational reference wave function was obtained by density matrix renormalisation group (DMRG) calculations with maxM = 4000 retained renormalised states. NEVPT2 and the MC-PDFT functionals used the same underlying reference wave function. All calculations employed the scalar-relativistic exact-two-component (X2C) Hamiltonian and the cc-pVTZ-DK basis set. Results were compared with the revised experimental reference curve obtained by Larsson and colleagues from reassigned vibrational levels [26].

The methods differ substantially in their predictions of the well depth and the shape of the stretched region. Without a dynamic-correlation correction, the DMRG reference curve markedly underestimates the binding energy near the experimental equilibrium and fails to reproduce the deep well. Adding dynamic correlation through NEVPT2 produces a more reasonable binding well. However, NEVPT2 still underestimates the binding strength and gives energies that are too high around the shoulder and at longer bond lengths. MC-PDFT performance varies with the functional. tM06-L and tB4LYP yield wells that are too shallow. MC23 and MC25 give wells that are too deep and remain below the experimental reference throughout the stretched region.

MC26 and COF26 improve the balance between the well depth and the stretched-region profile. Both produce a well near the experimental equilibrium bond length and rise smoothly through the shoulder without an unphysical maximum. COF26 still predicts overly negative binding energies near equilibrium and around the shoulder, corresponding to overbinding relative to the experimental reference. As the bond length increases, however, it approaches the RKR curve and closely matches the experimental reference at approximately 3.0–3.25 Å. In this region, COF26 outperforms MC26 and NEVPT2. Among the methods compared, COF26 gives the best overall agreement with the experimental reference and the lowest root-mean-square error (RMSE), 2.57 kcal $mol^{-1}$. These results show that, with a CAS(12e,22o) DMRG reference wave function, COF26 provides a balanced description across the bond-length range shown. This range extends from the equilibrium region to stretched $Cr_2$ bonds.

## Performance on the GMTKN55 sub database

The GMTKN55 database compiled by Goerigk and colleagues[28] covers main-group thermochemistry (TC), kinetics and non-covalent interactions. Morgante and Peverati[29] reported that the GMTKN55 reference values are more accurate than those in MGCDB84. We therefore selected GMTKN55 to benchmark the performance of COF26 for general main-group chemistry.

The 1,505 entries in GMTKN55 can be divided into five subdatabases. These cover basic properties and reaction energies of small systems ("Small"), reaction and isomerisation energies of large systems ("Large"), barrier heights ("BH"), intermolecular non-covalent interactions ("Inter-NC") and intramolecular non-covalent interactions ("Intra-NC"). We excluded W4-11 because of size-consistency concerns associated with the active-space protocol used here for atomisation energies.

Goerigk and colleagues introduced the weighted total mean absolute deviation metrics WTMAD-1 and WTMAD-2 [28] to compare density functionals on GMTKN55. By both measures, COF26 ranks among the five best of 26 functionals in each category: Small, Large, BH, Inter-NC and Intra-NC. In many cases, COF26 outperforms double-hybrid and machine-learning functionals. For the Radical7 subdatabase, COF26 gives a WTMAD-2 of 2.81 kcal mol$^{-1}$. This value is lower than those of MC26 (4.03 kcal mol$^{-1}$), MC23 (4.30 kcal mol$^{-1}$), DM21 (4.72 kcal mol$^{-1}$) and MC25 (4.74 kcal mol$^{-1}$). Across the GMTKN55 subsets, COF26 gives a WTMAD-2 of 3.64 kcal mol$^{-1}$, compared with 4.68 kcal mol$^{-1}$ for MC25. Its overall error is lower than those of several double-hybrid functionals examined, but remains above those of DSD-BLYP-D3(BJ) and B2GPPLYP-D3(BJ).

COF26 performs particularly well in the Small category because multireference wave functions and densities were used for the small systems. With standard GMTKN55 weights, COF26 ranks first for the Small subset with a WTMAD-1 of 1.66 kcal mol$^{-1}$. Its WTMAD-2 of 2.41 kcal mol$^{-1}$ ranks first among the methods shown and is slightly below 2.42 for DSD-BLYP-D3(BJ) (Fig. 8b). The COF26 WTMAD-2 is lower than those of four other double-hybrid functionals. B2GPPLYP-D3(BJ), PWPB95-D3(BJ), MPW2PLYP-D3(BJ) and B2PLYP-D3(BJ) give values of 2.55, 3.09, 3.20 and 3.31 kcal mol$^{-1}$, respectively. COF26 also outperforms the machine-learning functional DM21 at 2.59 kcal mol$^{-1}$ and slightly outperforms Skala 1.1 at 2.45 kcal mol$^{-1}$.

The advantage of COF26 is especially pronounced for the ALKBDE10 bond-dissociation energy subset. Its MUE of 1.88 kcal mol$^{-1}$ is lower than 4.22 and 3.79 kcal mol$^{-1}$ for the machine-learning functionals DM21 and Skala 1.1, respectively. It is also below the 2.78–3.23 kcal mol$^{-1}$ range of the five double-hybrid functionals compared. For the G21EA electron-affinity subset, COF26 gives an MUE of 1.14 kcal mol$^{-1}$. This value is lower than 1.34 kcal mol$^{-1}$ for DM21 and the 1.29–1.76 kcal mol$^{-1}$ range of the five double hybrids. It is close to 1.11 kcal mol$^{-1}$ for Skala 1.1. For the difficult-case subset DC13, the COF26 MUE of 2.74 kcal mol$^{-1}$ is also below 4.23 kcal mol$^{-1}$ for DM21. It is lower than the 3.71–6.77 kcal mol$^{-1}$ range of the five double hybrids, whereas Skala 1.1 gives a smaller error of 2.20 kcal mol$^{-1}$. The TAUT15 tautomerisation-energy subset shows a different ordering. The COF26 MUE of 0.52 kcal mol$^{-1}$ is below 0.54 kcal mol$^{-1}$ for B2GPPLYP-D3(BJ) and the 0.62–0.72 kcal mol$^{-1}$ range of three other double hybrids. However, DSD-BLYP-D3(BJ), DM21 and Skala 1.1 give MUEs of 0.46, 0.45 and 0.32 kcal mol$^{-1}$, respectively.

## NCIA non-covalent interaction datasets

Accurate prediction of non-covalent interactions requires a functional to balance attractive and repulsive effects across chemical environments and intermolecular separations. The Non-Covalent Interactions Atlas (NCIA) benchmark datasets developed by Řezáč and colleagues provide reference interaction energies from systematic sampling of intermolecular separations. The subsets selected here cover four classes: London dispersion [30], hydrogen bonding [31,32], short-range repulsion [33] and σ-hole interactions [34].

Figure 9 compares the overall errors and category decomposition of the functionals on NCIA. After weighting by the number of reactions, COF26 gives an overall MUE of 0.22 kcal mol$^{-1}$. This value is lower than 0.24, 0.25 and 0.30 kcal mol$^{-1}$ for MC26, MC25 and MC23, respectively. The errors are approximately 12% and 27% lower than those of MC25 and MC23. MC26 also improves overall accuracy while retaining the same analytical form as MC25.

The overall accuracy of COF26 is comparable to that of the best-performing KS-DFT functionals examined. Its overall MUE is close to that of ωB97X-D3(BJ), with both giving 0.22 kcal mol$^{-1}$. It is also below 0.24 and 0.25 kcal mol$^{-1}$ for ωB97X-V and B97M-V, respectively. The double-hybrid functional DSD-BLYP-D3 gives the lowest overall MUE of 0.18 kcal mol$^{-1}$. COF26 therefore provides competitive accuracy for the non-covalent interaction energies covered by NCIA.

Category-level analysis shows that COF26 improves upon MC26, MC25 and MC23 across all four interaction classes. Its within-category MUEs for London dispersion, hydrogen bonding, short-range repulsion and σ-hole interactions are 0.21, 0.22, 0.09 and 0.36 kcal mol$^{-1}$, respectively. For short-range repulsion, its MUE is lower than 0.12 and 0.14 kcal mol$^{-1}$ for DSD-BLYP-D3 and ωB97X-D3(BJ). This result indicates an advantage for COF26 in predicting the short-range contact energies examined. For σ-hole interactions, the within-category MUE decreases from 0.53 kcal mol$^{-1}$ with MC23 to 0.36 kcal mol$^{-1}$ with COF26. It nevertheless remains above 0.29 kcal mol$^{-1}$ for DSD-BLYP-D3. For London dispersion and hydrogen bonding, the accuracy of COF26 approaches that of ωB97X-D3(BJ), which gives MUEs of 0.20 and 0.22 kcal mol$^{-1}$. DSD-BLYP-D3 gives lower errors of 0.17 and 0.16 kcal mol$^{-1}$ for these two classes.

## Performance on subsets of the GSCDB137 database

Figure 10 summarises the normalised error ratio (NER) by category for 69 subsets of GSCDB137. The categories are barrier heights (BH), electric-field response (EF), vibrational frequencies (FREQ), isomerisation energies (ISO), intramolecular non-covalent interactions (INC), non-covalent interactions (NC), thermochemistry (TC) and transition-metal chemistry (TM). The figure also reports the overall mean. NER follows the GSCDB137 definition, in which each error is divided by a dataset-specific baseline: the mean of the second-, third- and fourth-lowest hybrid-functional errors in the source study [35].

We compared COF26 and MC26 with the modern functionals $\omega$B97M-V and CF22D, the on-top functionals MC23 and MC25, and the conventional hybrid functionals PBE0 and B3LYP. The reference results for ωB97M-V, CF22D, PBE0-D4 and B3LYP-D4 were taken from the GSCDB137 study of Liang and Head-Gordon. Most reference wave functions for the MC-PDFT calculations were single-configuration wave functions. CASSCF calculations were performed only for a small number of datasets with multireference character. The corresponding active-space choices are provided in Supplementary Data 2. PBE0 and B3LYP were augmented with D4 dispersion corrections using the $\omega$B97M-D4 parameters to describe non-covalent interactions. CF22D used its own fixed accompanying dispersion term [36].

COF26 gives the lowest overall mean NER of 1.10, an improvement of approximately 12% over CF22D value of 1.25. It also substantially outperforms the other on-top functionals MC23, MC25 and MC26. $\omega$B97M-V and CF22D rank second and third, respectively. Relative to $\omega$B97M-V and CF22D, the largest improvements from COF26 occur for ISO and TM. COF26 also leads in the INC and BH categories and is more accurate for intermolecular non-covalent interactions than the other on-top functionals. Across the 12 NC subsets, COF26 lowers the mean NER by approximately 5%, from 1.35 for MC26 to 1.28. Its value is also below 1.49, 1.94 and 2.45 for CF22D, B3LYP-D4 and PBE0-D4, respectively. $\omega$B97M-V ranks first with a mean NER of 1.01.

The category-mean NER of COF26 for TC is 1.05, ranking second among eight functionals behind only CF22D. Its category-mean NER for electric-field response is 1.75, ranking fifth among eight functionals. This value is lower than 1.78, 1.84 and 1.94 for $\omega$B97M-V, CF22D and B3LYP-D4, respectively. Relative to MC26, the category-mean NER of COF26 increases by approximately 11%. COF26 ranks sixth in the FREQ category, whereas MC26 ranks third and fifth in these two categories, respectively. Detailed results for each dataset are provided in the Supplementary Information.

# Discussion

Density functional theory (DFT) is the most widely used electronic-structure method. However, current KS-DFT functionals are optimized only for single-reference systems, and few functionals accurately predict all properties required in complex chemical applications. Here, agent-driven supervised learning systematically optimized an on-top functional. We introduced a nonseparable exchange-correlation term, constructed a large benchmark database and developed COF26. Its performance on GMTKN55 subsets is comparable to that of the best double-hybrid functionals examined. It also quantitatively describes the equilibrium region, shoulder behaviour of the $Cr_2$ potential-energy curve. For single-reference systems, densities obtained from RHF or KS-DFT/PBE offer a balance between accuracy and efficiency. As shown in the Results, COF26 gives the best overall accuracy across broad chemical space. The benchmarks cover main-group thermochemistry, multireference systems, transition-metal energetics and excited states. These results support applications of COF26 to diverse bonding and non-covalent interactions in main-group and transition-metal compounds. Such applications may include studies of catalysis, functional materials, biochemistry and environmental chemistry.

COF26 nevertheless has limitations. For multireference systems, the accuracy of MC-PDFT fundamentally depends on the quality of the underlying multiconfigurational reference wave function. CASSCF and DMRG calculations remain substantially more expensive than KS-DFT, and their scaling in large systems presents a practical bottleneck. Fragmented CAS methods such as LASSCF can partly address this issue [37]. This limitation is not unique to MC-PDFT. It applies to all post-processing methods based on multireference wave functions, including CASPT2, NEVPT2 and internally contracted multireference configuration interaction. Moreover, the current COF26 energy descriptors are constructed entirely from local variables. They do not explicitly describe long-range dispersion or non-local exchange-correlation effects, resulting in a systematic limitation where long-range interactions dominate.

The FunctionalAgent could be extended to explore alternative functional representations. Their energies could depend not only on local density variables, but also on environmental descriptors from non-local density convolutions or molecular representations extracted by equivariant neural networks. Such non-local extensions of on-top functionals could systematically incorporate physical effects that local approximations cannot describe. Continued advances in density matrix renormalisation group methods and LASSCF are expanding the accessible active spaces, thereby broadening the systems amenable to MC-PDFT. High-accuracy multireference benchmark data may eventually extend beyond current coverage of main-group and 3*d* transition-metal chemistry. As these datasets grow, MC-PDFT functionals may show systematic gains in accuracy similar to those already observed for deep-learning functionals in KS-DFT. Continued development of MC-PDFT, together with systematic functional-design strategies such as agent-driven optimisation, provides a sound methodological basis for progressively improving quantitative descriptions of these challenging chemical problems.

# Methods

## FunctionalAgent: an agent-driven optimisation workflow

FunctionalAgent comprises an orchestration layer and three specialised subagents (Fig. 1). The orchestration agent acts as a central coordinator. It converts researcher-defined optimisation objectives into structured task plans, dispatches requests to subagents, tracks execution and updates the exploration strategy from aggregated results. Long-running optimisation tasks may span hundreds of molecular systems. The orchestration agent therefore communicates with subagents using only bounded, decision-relevant context to prevent context-window overflow. This approach also ensures that each subagent receives only the information required for its current task. Detailed computational files, including input geometries, wave-function checkpoints and descriptor tables, are exchanged through a shared file bus. Before full execution, the orchestration agent enters planning mode and generates a detailed workflow for human review and approval.

## Active-space selection and input generation

Accurate multireference calculations require appropriate active spaces. Manual specification becomes impractical, however, when a benchmark suite contains hundreds of molecular systems with distinct bonding patterns. The Active-Space Input Agent automates this bottleneck. It receives standardised data entries and reorganises molecular coordinates and computational tasks by molecular system and reaction type. The agent first performs Hartree--Fock calculations to obtain initial orbital information. It then combines chemically motivated rules with automated schemes such as AVAS [38] and autoCAS [39] to specify active spaces. For example, single-configuration wave functions are assigned to single-reference systems, whereas valence π orbitals and lone-pair orbitals are selected for organic chromophores. The agent chooses an appropriate scheme from the electronic-structure characteristics of each system. The specified active spaces are converted into input files for PySCF [40] or OpenMolcas [41]. By combining domain heuristics with orbital-analysis methods, this module balances coverage across chemically diverse datasets. It also reduces the risk of omitting strongly correlated orbitals in individual systems.

## Quantum-chemical calculations and descriptor generation

Once the active spaces are defined, large-scale multireference calculations must be executed, monitored and post-processed reproducibly. The QChem Calculation Agent manages this pipeline from end to end. It submits input files to computational clusters through HPC connection and file-transfer interfaces. The agent invokes CASSCF [42] or related multireference calculations, monitors job status and retrieves output files and wave-function checkpoints. It then parses the results to extract descriptors related to the electron density ρ and on-top pair density Π. As described below for the functional forms, these descriptors are organised into feature representations. They include conventional CASSCF energy terms and linear on-top functional energy contributions built using the Libxc exchange-correlation functional library [43]. By automating the submission, monitoring and extraction of hundreds of concurrent jobs, this module removes a manual bottleneck that commonly limits the scale of functional development. It also ensures consistent descriptor definitions across all molecular systems.

## Functional optimisation

The central challenge of functional development is to search a high-dimensional parameter space while satisfying training accuracy, out-of-sample generalisation and robustness across datasets. These objectives often conflict when optimised independently. Training-set selection and weight determination in the large MCDDB26 database produce an optimisation space $10^{31}$ times larger than in the previous MC23 or MC25 optimisations. The Functional Optimization Agent addresses this problem by combining a performance-triggered supervised-learning cycle with Bayesian optimisation.

The overall optimisation has two components. A constrained inner optimiser restricts fitting to a predefined functional form and optimises only its learnable linear parameters. It uses the sequential least-squares programming (SLSQP) algorithm implemented in SciPy and JAXopt [44].

The performance-triggered supervised-learning cycle comprises five steps:

1. Dataset reweighting and regularisation refinement. FunctionalAgent uses Bayesian optimisation over joint metric performance to rebalance dataset weights. This process provides a systematic, data-driven exploration of the weight space.

2. Extended testing and performance diagnosis. The current model is evaluated on external datasets not used for training. Diagnostics identify underperforming system types at both the category level, such as transition-metal thermochemistry and barrier heights, and the individual-dataset level. Performance is quantified using several metrics, including root-mean-square error (RMSE), mean unsigned error (MUE) and dataset-level rank. The agent integrates these metrics into a comprehensive diagnostic assessment.

3. **Data augmentation and training-set expansion.** External tests may reveal persistent systematic weaknesses in specific chemical categories that reweighting alone cannot resolve. Challenging datasets or chemically related problem classes are then added to the training set, providing the optimiser with sufficient coverage of the problematic regions.

4. Model retraining and external validation. The model is retrained on the updated training set. Its performance on both the training and external test sets is compared with the current best model and reference functionals such as MC23 [18] and MC25 [19].

5. Candidate-model selection. A candidate replaces the current best model only if it improves global accuracy metrics, such as overall MUE across all datasets, without substantially degrading any key dataset. This joint criterion prevents optimisation of aggregate metrics from concealing local performance degradation in chemically important subsets.

The closed-loop optimisation terminates when the model reaches the user-defined accuracy target or fails to improve meaningfully over the current best model for several successive iterations. It also terminates when both training and external metrics plateau, indicating that the available training data cannot support further improvement.

## Functional design

Previous studies have described MC-PDFT [12], L-PDFT [45], hybrid functionals [46], meta functionals [18] and their application in L-PDFT [47] in detail. Here, we provide only a brief overview of the main theoretical framework. We first consider singlet MC-PDFT calculations. The hybrid MC-PDFT energy is defined as

$$E_{\mathrm{MCPDFT}} = E_{\mathrm{class}} + (X/100)E_{\mathrm{MC,XC}} + [1 - (X/100)]E_{\mathrm{ot}}$$

where $E_{\mathrm{class}}$ is the classical energy computed from the MCSCF wave function. $E_{\mathrm{MC,XC}}$ is the exchange-correlation, or nonclassical, energy of the MCSCF wave function. The parameter $X$ is expressed as a percentage, and $E_{\mathrm{ot}}$ is the nonclassical energy computed using an on-top functional.

For multistate L-PDFT calculations of excited-state systems, state energies are obtained by diagonalising the model-space effective Hamiltonian matrix in the basis of SA-CASSCF eigenvectors. The model-space dimension equals the number of states included in the state averaging of the SA-CASSCF calculation. The effective Hamiltonian operator is constructed as a first-order Taylor expansion around a zeroth-order reference energy. This reference involves the state-averaged one- and two-electron density matrices obtained from SA-CASSCF. Full details are available in previous studies [45,47].

The optimised hybrid meta-on-top functional MC26 uses the same analytical form as MC25. Specifically, it uses translated M06-L and includes a fraction $X$ of the CASSCF wave-function energy within the hybrid MC-PDFT framework. All 38 linear parameters in tM06-L and $X$ are reoptimised simultaneously.

The form of the new COF26 functional is based on the previously successful MN15-L and M06L functionals [48]. COF26 additionally introduces a nonseparable exchange-correlation energy. The on-top energy is written as a linear combination of the nonseparable exchange-correlation energy $E_{\mathrm{nxc}}$, exchange energy $E_x$ and additional correlation energy $E_c$:

$$E_{\mathrm{ot}} = E_{\mathrm{nxc}} + E_x + E_c.$$

where

$$E_{\mathrm{nxc}} = \int \mathrm{d}\mathbf{r} \sum_{\sigma=\alpha}^{\beta} \rho_\sigma \left\{ \varepsilon_{x\sigma}^{\mathrm{LSDA}}(\rho_\sigma) \sum_{i=0}^{3} \sum_{j=0}^{3-i} \sum_{k=0}^{5-i-j} a_{ijk} \{v_{x\sigma}(\rho_\sigma)\}^i \{u_{x\sigma}(s_\sigma)\}^j \{w_\sigma(\rho_\sigma, \tau_\sigma)\}^k \right\},$$

$$E_x = \sum_\sigma \int \mathrm{d}\mathbf{r} \left[ F_{x\sigma}^{\mathrm{PBE}}(\rho_\sigma, \nabla\rho_\sigma) f(w_\sigma) + \varepsilon_{x\sigma}^{\mathrm{LSDA}} h_x(x_\sigma, z_\sigma) \right]$$

$$E_c = \int \mathrm{d}\mathbf{r}\, \rho \varepsilon_c^{\mathrm{LSDA}}(\rho_\alpha, \rho_\beta) \left( \sum_{i=0}^{11} b_i \{w(\rho, \tau)\}^i \right) + \int \mathrm{d}\mathbf{r}\, \rho H^{\mathrm{PBE}}(\rho_\alpha, \rho_\beta, s) \left( \sum_{i=0}^{11} c_i \{w(\rho, \tau)\}^i \right) + \int \mathrm{d}\mathbf{r}\, e_{\alpha\beta}^{\mathrm{UEG}} \left[ g_{\alpha\beta}(x_\alpha, x_\beta) + h_{\alpha\beta}(x_{\alpha\beta}, z_{\alpha\beta}) \right],$$

Here, $\rho_\alpha$ and $\rho_\beta$ are the translated spin-up and spin-down electron densities at spatial point $r$, and $\rho$ is their sum. $\tau_\alpha$ and $\tau_\beta$ are the spin-up and spin-down kinetic-energy densities, respectively. The functions $v_{x\sigma}$, $u_{x\sigma}$, $w_\sigma$, $\varepsilon_c^{\mathrm{LSDA}}$ and $H^{\mathrm{PBE}}$ are the same as those in the MN15-L functional. The functions $F_{x\sigma}^{\mathrm{PBE}}$, $\varepsilon_{x\sigma}^{\mathrm{LSDA}}$, $e_{\alpha\beta}^{\mathrm{UEG}}$, $h_x$, $f$, $g_{\alpha\beta}$ and $h_{\alpha\beta}$ are the same as those in the M06-L functional. We therefore do not redefine them here.

During functional-parameter optimisation, the first term of the loss function is defined as a weighted sum of the RMSEs for the new functional across the training datasets. FunctionalAgent automatically assigns dataset weights according to performance during training. The optimal parameters of MC26 and COF26 are obtained by minimising this objective. A regularisation term is added to suppress overfitting and ensure the stability of the fitted parameters. The total loss function is therefore

$$\mathcal{L} = \sum_d w_d R_d + w_{\mathrm{reg}} \sum_q p_q^2$$

where $w_d$ is the weight of dataset $d$, $R_d$ is its RMSE and $p_q$ denotes a linear parameter in the functional form. The quantity $w_{\mathrm{reg}}$ is the regularisation parameter. Functionals containing one or more large $p_q$ values are generally less smooth and therefore more susceptible to overfitting. FunctionalAgent iteratively determines $w_{\mathrm{reg}}$ during optimisation to maintain the generalisation performance of the functional.

## MCDDB26 database

We constructed a composite database named the Multi-Configuration Diverse Database 2026 (MCDDB26), which contains 150 datasets and 18,954 data points. All component datasets are summarised in Fig. 3 and described in detail in Supplementary Data 1. MCDDB26 was assembled from six source categories.

- Ground-state databases used to train MC23 and MC25 include Dataset 2 (DS2) from our previous work [49], which is based on Minnesota Database 2019 [50]. We also included Dataset 3 (DS3) from the MC25 study [19]. Each database contains several subdatabases with at least two data points. The MC25 PC-ED65 dataset, generated from seven potential-energy curves, was also included.

- The excitation-energy datasets used to train MC25 include the TM-SpinSplitting10 and MG-SpinSplitting26 databases and vertical excitation energies from QUESTDB [19,51–54]. The accurate QUESTDB reference values were originally reported by Loos and colleagues in a series of studies. For all excitation-energy data, the excitation energies and descriptors were calculated using the same wave functions as in the MC25 study.

- The GMTKN55 database developed by Goerigk and colleagues covers general main-group thermochemistry, kinetics and non-covalent interactions [28].

- We constructed the Active Space Robustness Dataset (ASRD) to train and evaluate the robustness of functional performance to active-space selection. ASRD includes the Large-CAS* dataset, generated by expanding the active spaces of selected systems. It also includes MGBB, a main-group bond-breaking dataset derived from several potential-energy curves [55–58]. Finally, it contains the Cr2-dimer-12o and Cr2-dimer-22o datasets, whose reference curves are based on the recently fitted experimental potential-energy curve of Larsson and colleagues [26].

- Selected subsets of the Gold-Standard Chemical Database 137 (GSCDB137) [59].

- Selected data from the Non-Covalent Interactions Atlas (NCIA) benchmark series developed by Řezáč and colleagues span four classes of non-covalent interactions. London dispersion is represented by D442×10 [30]. Hydrogen bonding is represented by HB375×10 [31] and HB300SPX×10 for extended chemical space [32]. Short-range repulsion is represented by R739×5 [33], and σ-hole interactions by SH250×10 [34]. These subsets provide interaction energies sampled across intermolecular separations. They assess functional accuracy across contact distances and interaction types while extending MCDDB26 coverage of intermolecular non-covalent interactions.

### Additional data and references

Additional data from this study and additional references are provided in the Supplementary Information.

## Corresponding Author

*E-mail: xiaohe@phy.ecnu.edu.cn (X.H.)

## ORCID

Yuhao Chen: 0009-0009-8054-0923

Dayou Zhang: 0000-0002-5393-657X

Donald G. Truhlar: 0000-0002-7742-7294

Xiao He: 0000-0002-4199-8175

## Acknowledgments

X.H. was supported by the National Natural Science Foundation of China (Grant Nos. 22633005, 92477103 and 22273023), the Shanghai Municipal Science and Technology Commission (Grant No. 25511102400), the Shanghai Municipal Commission of Economy and Informatization (Grant No. 2026-GZL-RGZN-01015), the Advanced Materials-National Science and Technology Major Project (Grant No. 2026ZD0623702), the Shanghai Frontiers Science Center of Molecule Intelligent Syntheses, and the Fundamental Research Funds for the Central Universities. We acknowledge the Supercomputer Center of East China Normal University (ECNU Multifunctional Platform for Innovation 001) for providing computational resources.

## Author contributions

Y.C., D.Z., D.G.T. and X.H. performed the research. Y.C. and X.H. analysed the data. Y.C., D.Z., D.G.T. and X.H. wrote the paper.

## Data availability

Detailed descriptions of the component datasets in MCDDB26 are provided in Supplementary Data 1. The optimised parameters of the COF26 and MC26 functionals are provided in the Supplementary Information. Wave-function and energy-descriptor results for all datasets discussed here are available from Zenodo. Source data for Fig. 4 and Fig. 8 are provided with this paper.

# Code availability

Implementations of the COF26 and MC26 functionals are available in the master branch of the PySCF repository: https://github.com/chen-yu-hao/pyscf.git.

# Competing interests

The authors declare no competing interests.

Figures and figure legends

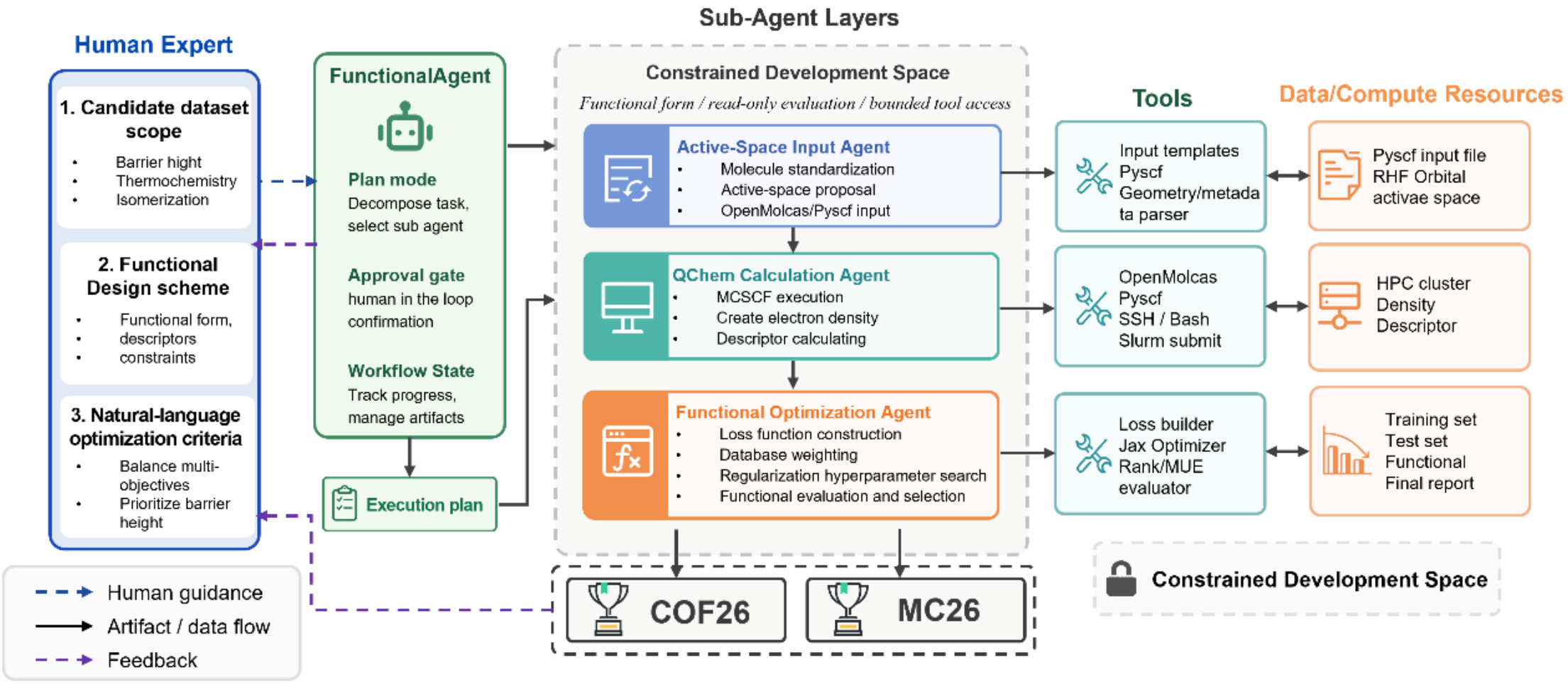

**Figure 1 | Constrained LLM-assisted agentic optimisation framework for developing MC26 and COF26.** The agent supports fitting MC-PDFT functionals within a constrained chemical space. Researchers define the candidate-dataset scope, analytical functional form and optimisation criteria. FunctionalAgent then performs active-space and input generation, quantum-chemical calculations and descriptor generation, and functional optimisation and evaluation.

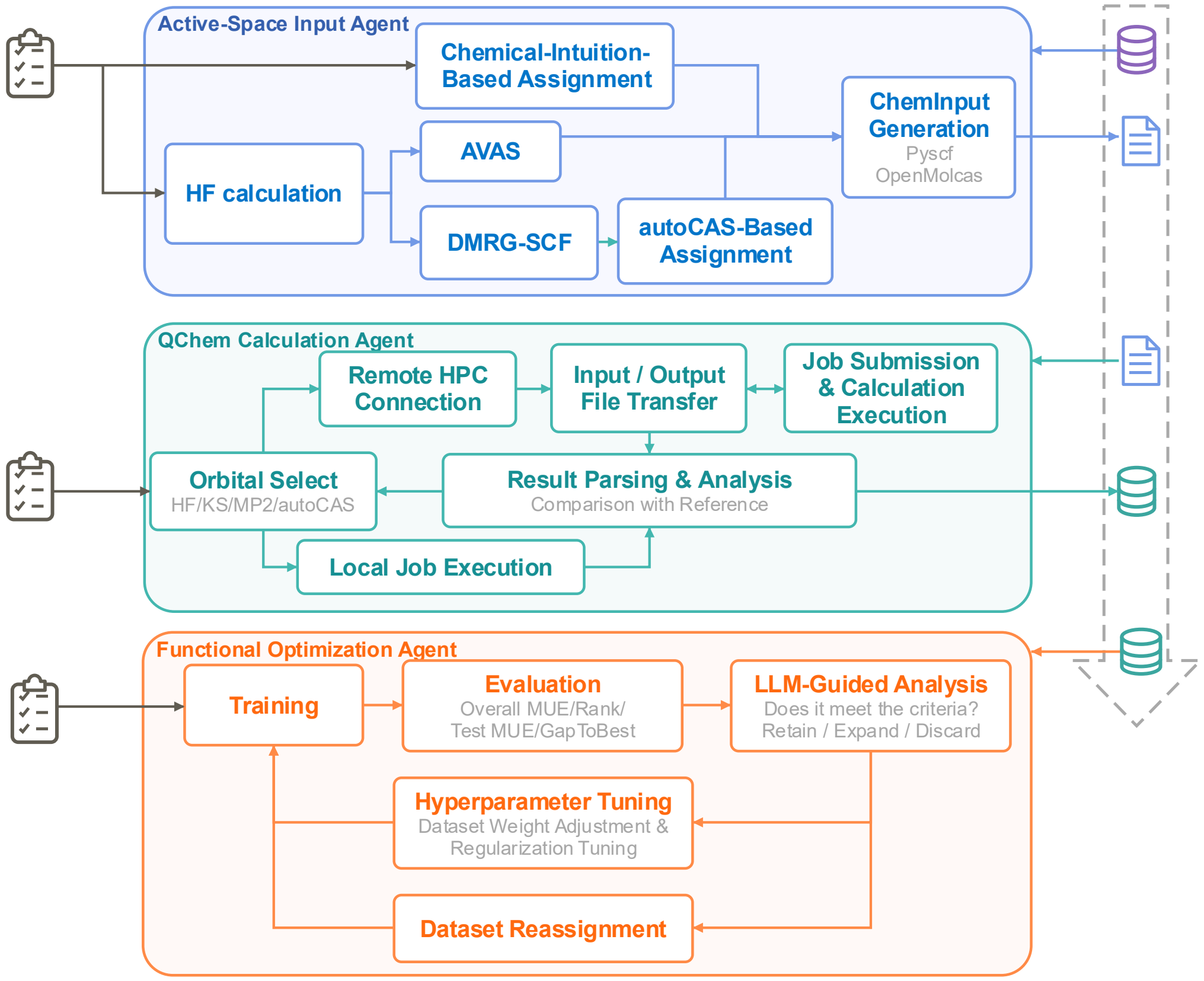

**Figure 2 | LLM-assisted workflows for constructing and refining quantum-chemical benchmark datasets. a,** Example workflow executed by the Active-Space Input Agent to define, parse and generate the active-space input files required for functional optimisation. **b,** Example workflow for MCSCF calculations and descriptor generation. **c,** Example performance-triggered supervised-learning cycle for iterative functional optimisation guided by the Functional Optimization Agent.

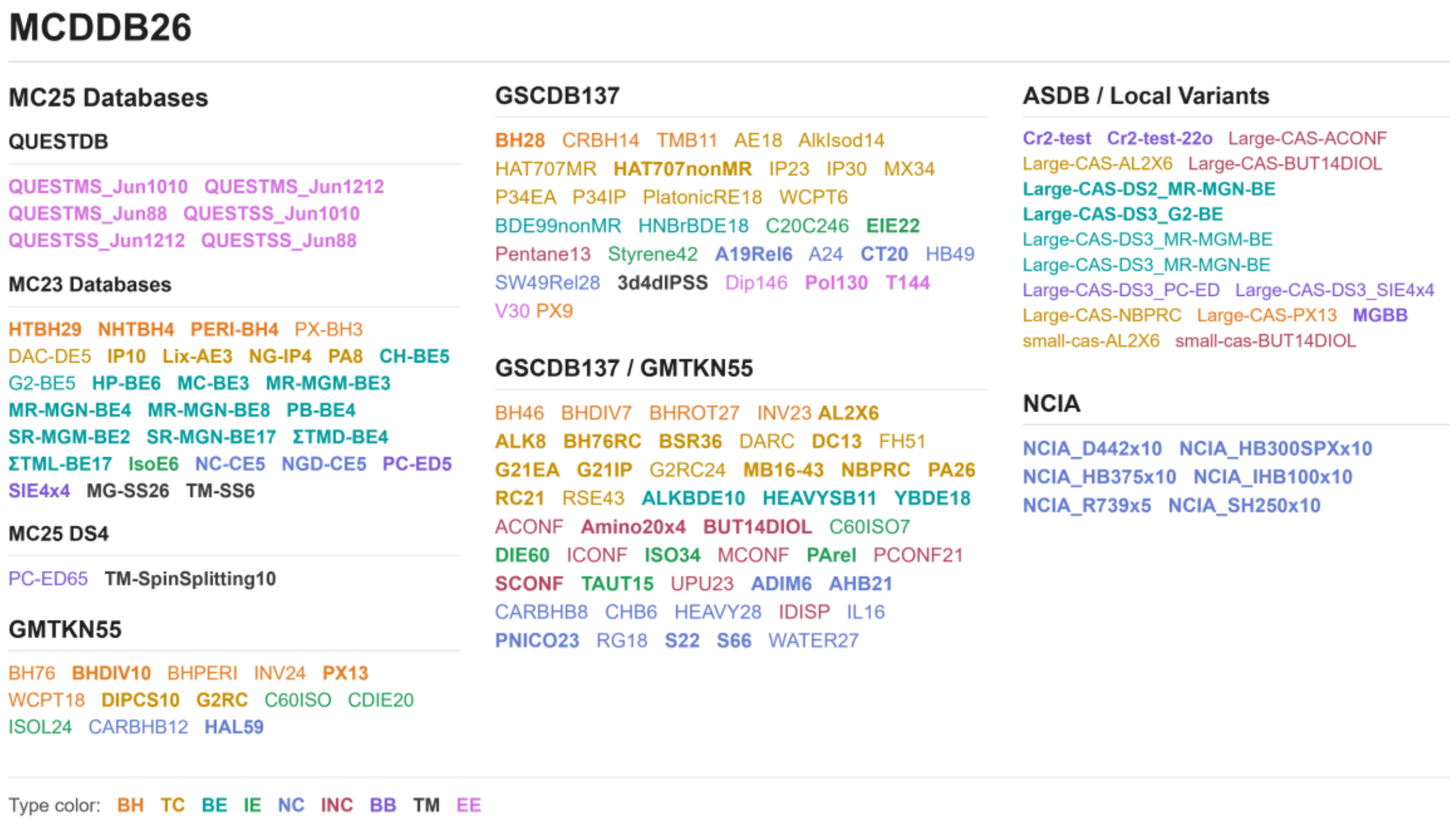

**Figure 3 | Composition of the MCDDB26 database.** Overview of the benchmark subsets used to develop and evaluate MC-PDFT on-top functionals in MCDDB26. Bold text denotes datasets in the training set (see Supplementary Data 1 for details). Each label represents a dataset or data subset, and colours denote the corresponding chemical categories. These are barrier heights (BH), thermochemistry (TC), bond energies (BE), isomerisation energies (IE), non-covalent interactions (NC), intramolecular non-covalent interactions (INC), bond breaking (BB), transition-metal energetics (TM) and excited-state energies (EE). MCDDB26 covers chemically and electronically diverse systems. These include main-group thermochemistry and kinetics, conformational and non-covalent interactions, transition-metal and spin-splitting energetics, QUESTDB excited states, and large-active-space bond-dissociation benchmarks such as $Cr_2$. This broad coverage provides a systematic basis for training, optimising and externally validating transferable on-top functionals.

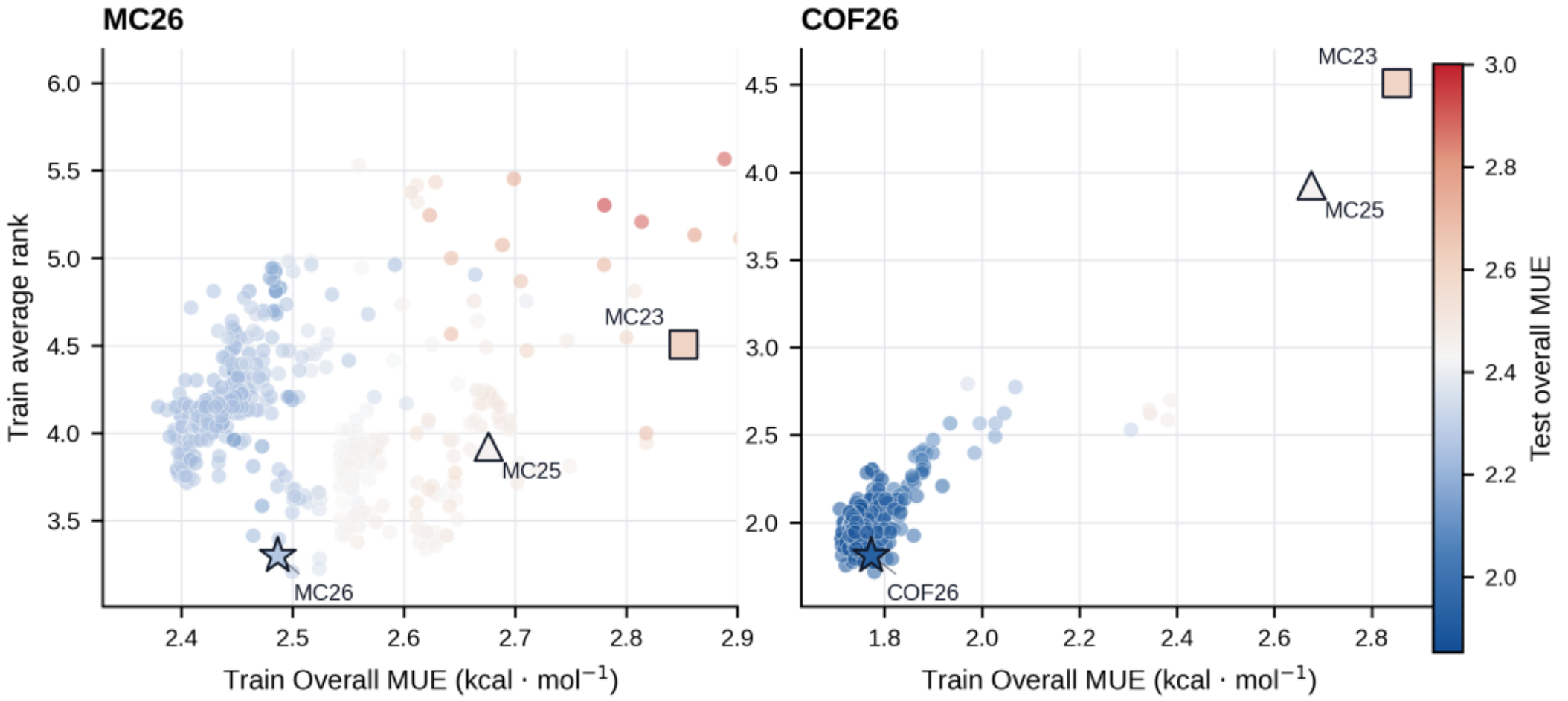

**Figure 4 | Sampling space during performance-triggered supervised-learning optimisation of the MC26 and COF26 functionals.** The scatter plot relates the overall training-set MUE to the mean training-set rank along the optimisation trajectories leading to MC26 and COF26. The colour of each point represents the corresponding overall test-set MUE. Labelled markers denote the reference and optimised functionals MC23, MC25, MC26 and COF26. MC26 and COF26 occupy low-error regions of the sampling space and have low test-set MUEs. They therefore improve generalisation while retaining strong training-set performance.

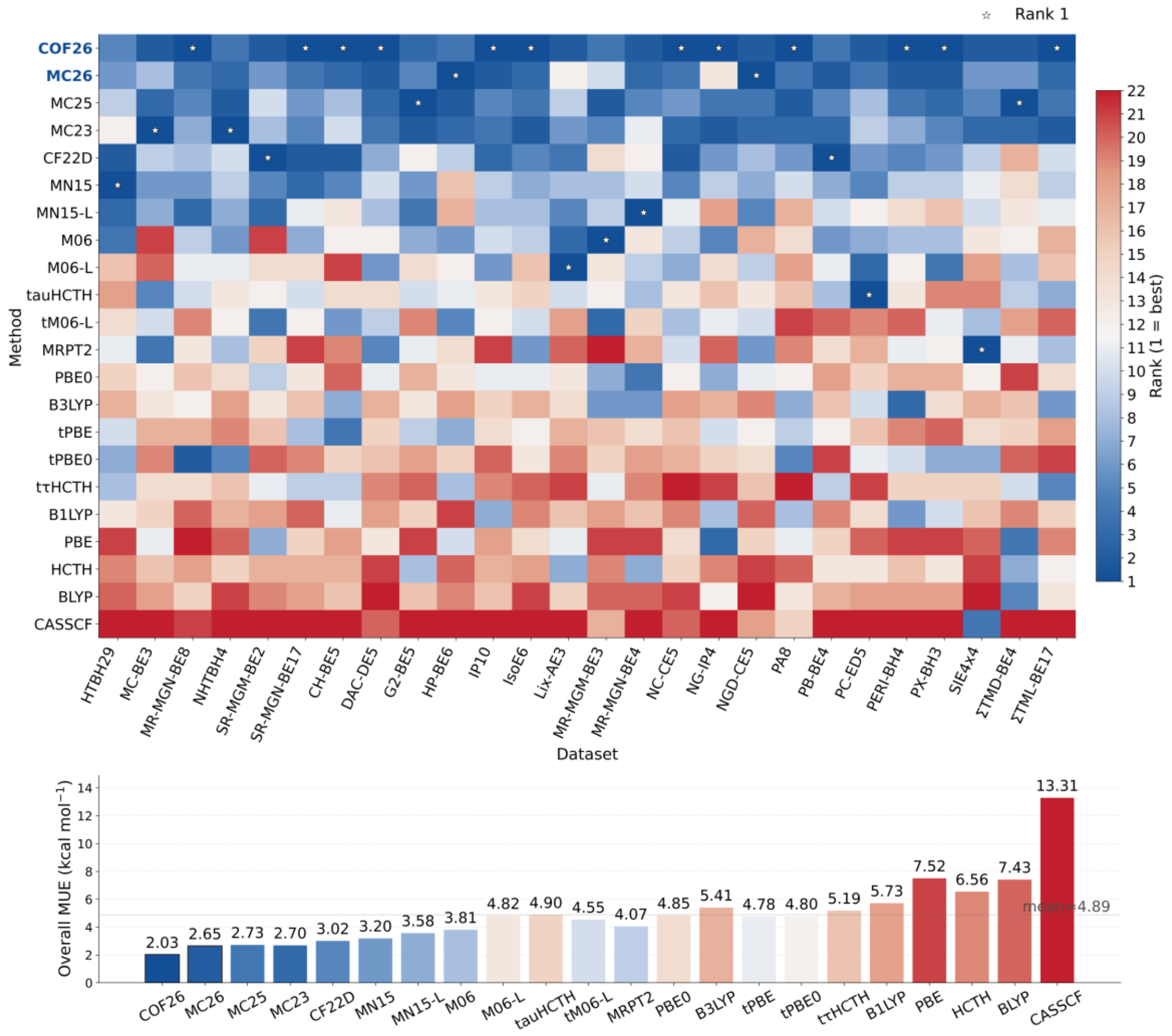

**Figure 5 | Per-dataset ranks and overall MUEs of COF26, MC26 and reference methods on selected ground-state training sets.** The upper panel shows the ranks of COF26, MC26 and the reference methods on comparable training subsets. Rank 1 indicates the best-performing method on a subset. Colors from blue to red denote increasing ranks, and white stars mark the top-ranked method for each subset. The lower panel summarizes the overall MUE of each method across the ground-state training subsets. Methods are ordered by mean rank. COF26, developed in this study, gives the best overall mean rank among all methods examined.

| database | COF26 | MC26 | MRPT2 | MC25 | tPBE0 | MC23 | MC25L | tPBE | tτ-HCTH | tM06-L | CASSCF |
|---|---|---|---|---|---|---|---|---|---|---|---|
| same-spin MUE in eV | | | | | | | | | | | |
| QUEST-Jun88-SameSpin28 | 0.14 | 0.16 | 0.14 * | 0.17 | 0.16 | 0.19 | 0.22 | 0.25 | 0.23 | 0.29 | 0.32 |
| QUEST-Jun1010-SameSpin28 | 0.15 | 0.16 | 0.11 * | 0.16 | 0.17 | 0.18 | 0.22 | 0.24 | 0.22 | 0.28 | 0.29 |
| QUEST-Jun1212-SameSpin31 | 0.15 | 0.16 | 0.11 * | 0.17 | 0.17 | 0.19 | 0.19 | 0.23 | 0.21 | 0.27 | 0.28 |
| spin-change MUE in eV | | | | | | | | | | | |
| QUEST-Jun88-SpinChange36 | 0.12 * | 0.12 | 0.15 | 0.13 | 0.13 | 0.16 | 0.21 | 0.18 | 0.24 | 0.26 | 0.21 |
| QUEST-Jun1010-SpinChange38 | 0.08 * | 0.08 | 0.09 | 0.08 | 0.09 | 0.12 | 0.14 | 0.13 | 0.16 | 0.20 | 0.23 |
| QUEST-Jun1212-SpinChange40 | 0.08 * | 0.09 | 0.10 | 0.09 | 0.09 | 0.13 | 0.14 | 0.13 | 0.16 | 0.19 | 0.26 |
| TM-SpinSplitting10 | 0.22 | 0.20 | 0.55 | 0.11 | 0.23 | 0.16 | 0.09 * | 0.25 | 0.29 | 0.41 | 0.97 |
| MG-SpinSplitting26 | 0.18 | 0.17 | 0.14 * | 0.18 | 0.16 | 0.20 | 0.21 | 0.20 | 0.21 | 0.21 | 0.27 |
| averaged by database in eV | | | | | | | | | | | |
| same-spin average MUE | 0.15 | 0.16 | 0.12 * | 0.17 | 0.17 | 0.19 | 0.21 | 0.24 | 0.22 | 0.28 | 0.30 |
| spin-change/splitting average MUE | 0.14 | 0.13 | 0.21 | 0.12 * | 0.14 | 0.15 | 0.16 | 0.18 | 0.21 | 0.25 | 0.39 |
| average of the two subsets | 0.14 * | 0.15 | 0.16 | 0.14 | 0.15 | 0.17 | 0.18 | 0.21 | 0.22 | 0.27 | 0.34 |

Rank (1 = best): 1 2 3 4 5 6 7 8 9 10 11

**Figure 6 | MUEs and ranks of COF26, MC26 and reference methods on the excited-state training subsets.** Values in the table are mean unsigned errors (MUEs) in eV. The bottom three rows report the arithmetic mean MUE of the three same-spin subsets and the mean of the five spin-changing and spin-splitting subsets. They also report the equally weighted mean of these two category means. Cell colors indicate method ranks in each row, increasing from 1 to 11 from blue to red. An asterisk marks the method with the lowest MUE in that row.

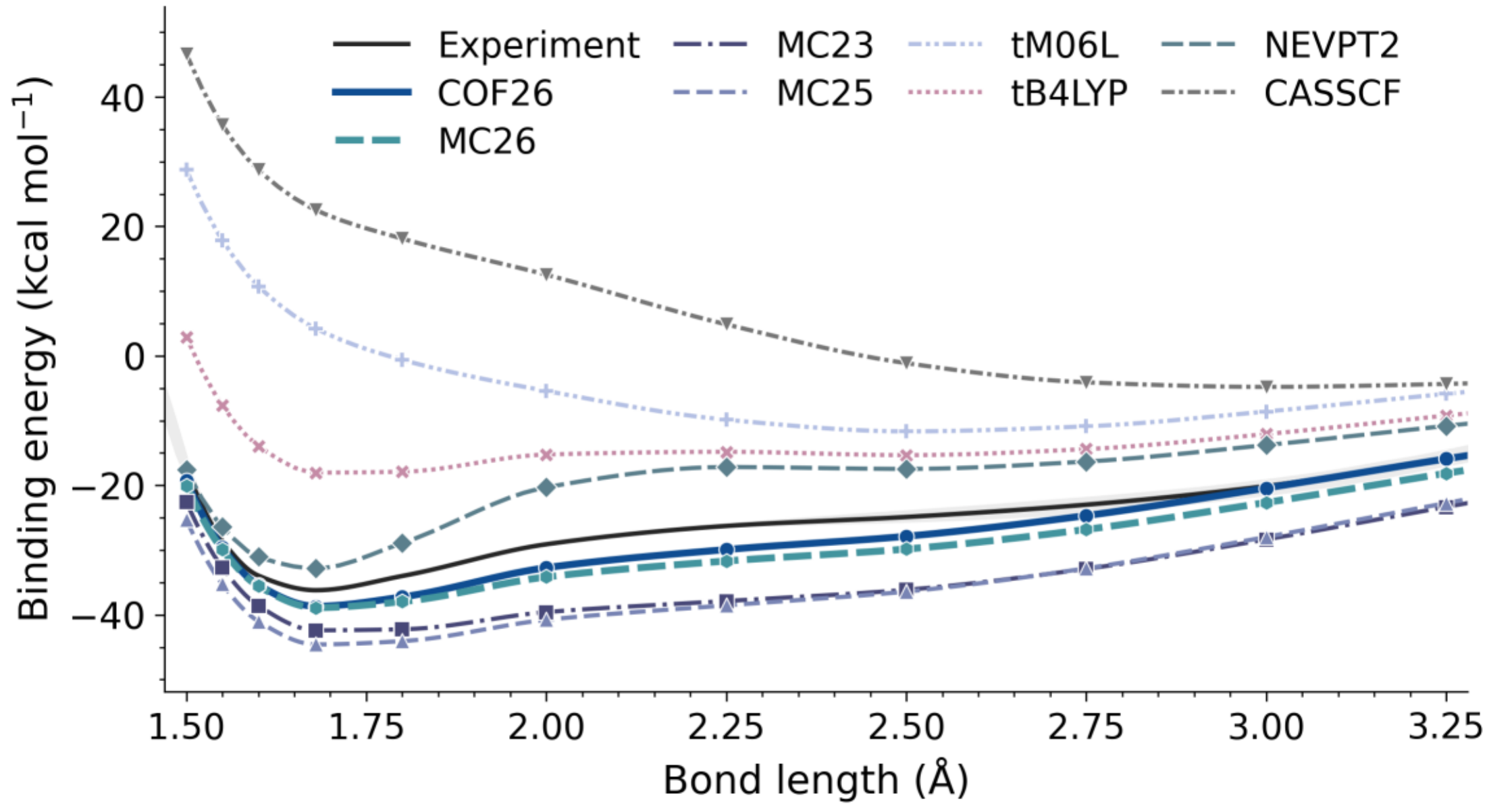

**Figure 7 | $Cr_2$ bond-dissociation curves calculated using different methods.** DMRG calculations used a CAS(12e,22o) active space with a maximum number of retained renormalized states of maxM = 4000. The NEVPT2 and MC-PDFT calculations used the same underlying multiconfigurational reference wave function. The experimental reference is an RKR

potential-energy curve fitted to vibrational-level spectroscopic data. All calculations used the scalar-relativistic exact-two-component (X2C) Hamiltonian and the cc-pVTZ-DK basis set.

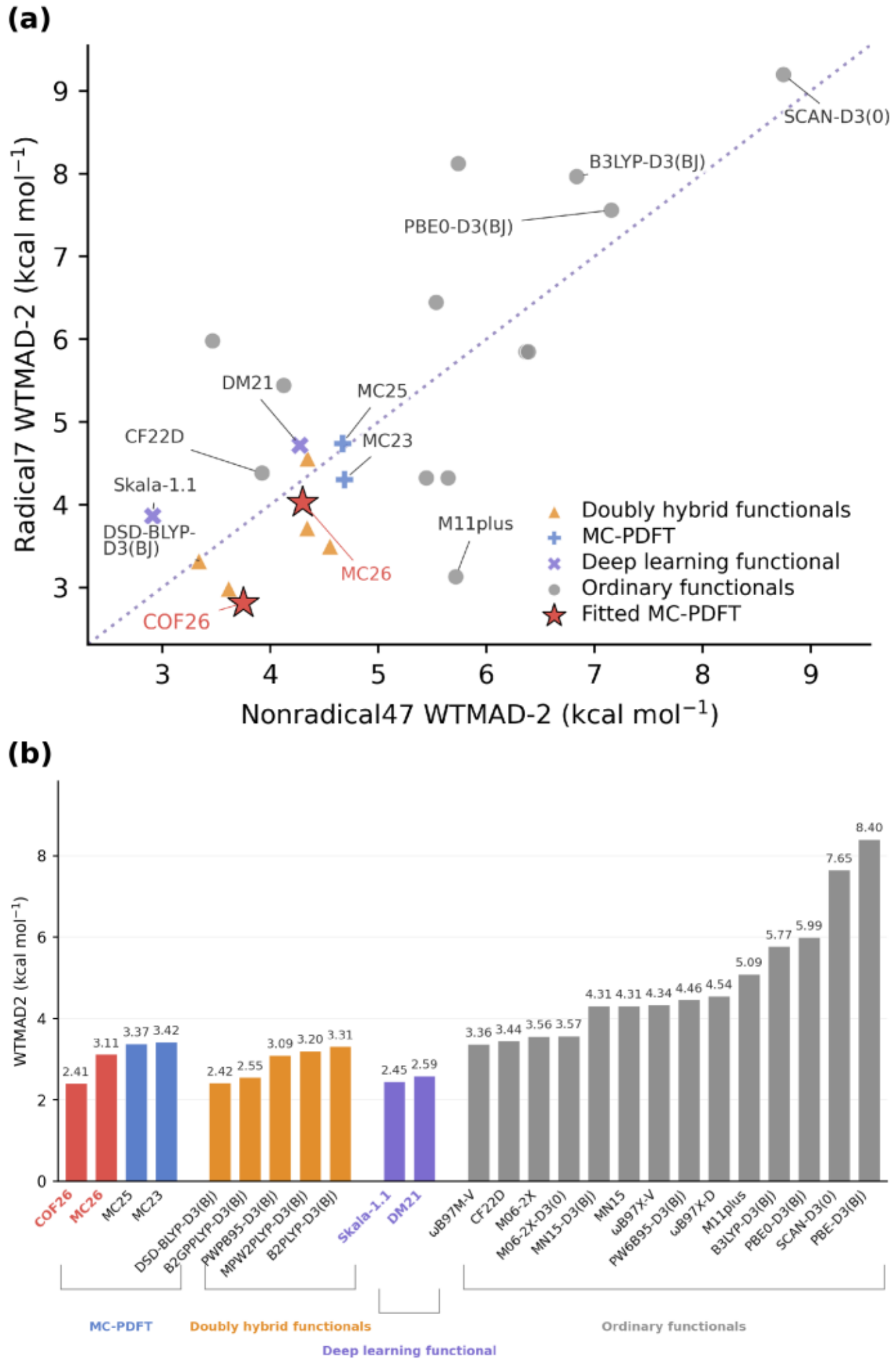

**Figure 8 | Benchmark performance of MC-PDFT functionals on GMTKN55 subsets. a,** WTMAD-2 of COF26 and selected functionals for the Radical7 subdatabase plotted against their WTMAD-2 for the Nonradical47 subdatabase. **b,** WTMAD-2 of the selected functionals for the Small subdatabase of GMTKN55. Lower values indicate smaller overall errors.

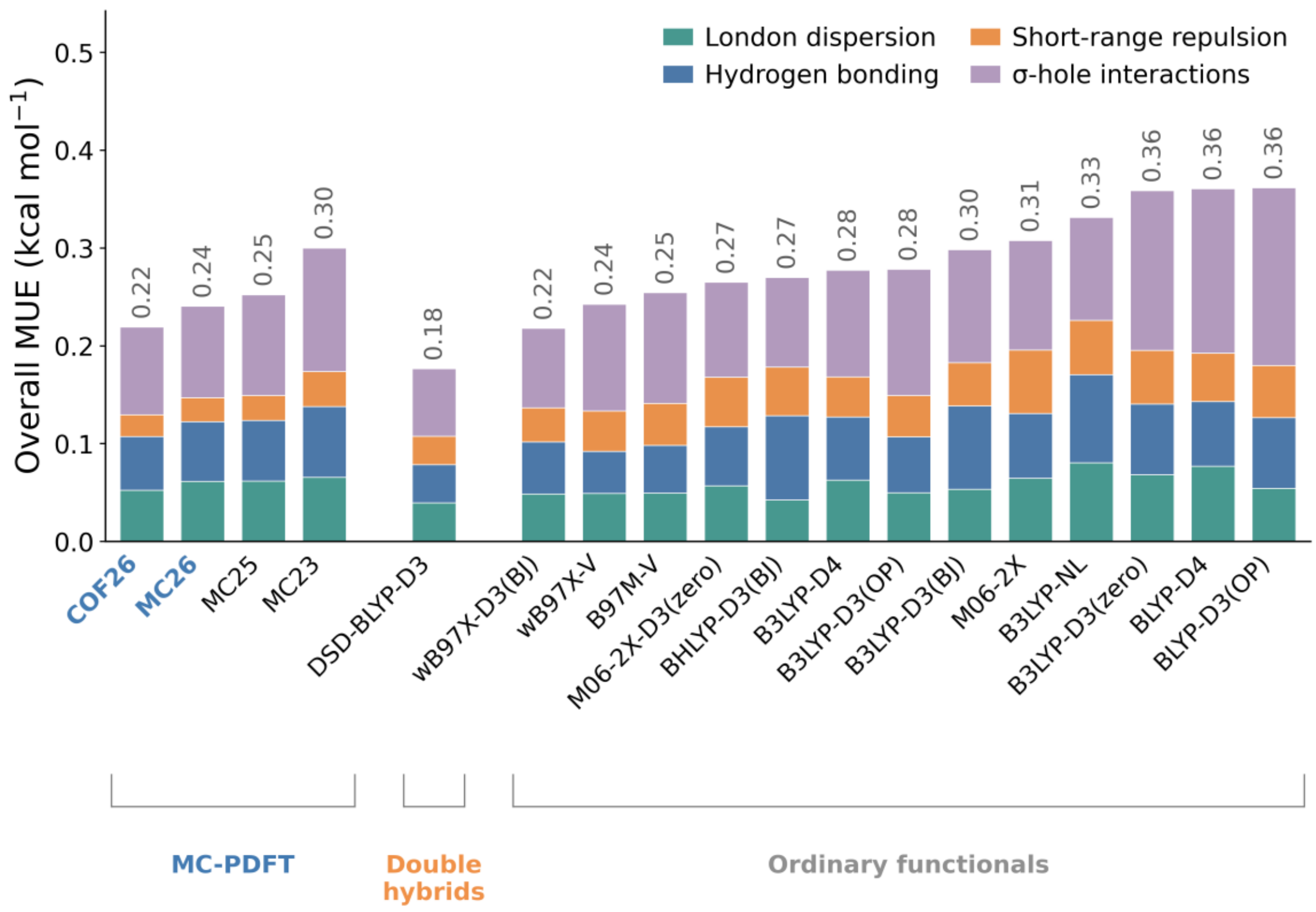

**Figure 9 | Reaction-count-weighted overall MUE and category decomposition for the NCIA datasets.** Bar heights represent the overall MUE for the NCIA datasets in kcal mol$^{-1}$. Stacked components show the contributions from London dispersion, hydrogen bonding, short-range repulsion and σ-hole interactions.

| Functional | Mean NER | NC | INC | TC | BH | EF | FREQ | ISO | TM |
|---|---|---|---|---|---|---|---|---|---|
| **COF26** | **1.10** | 1.28 | **1.01** | 1.05 | **0.89** | 1.75 | 1.31 | **1.04** | **0.89** |
| **MC26** | 1.28 | 1.35 | 1.97 | 1.13 | 0.95 | 1.57 | 1.16 | 1.15 | 0.95 |
| MC23 | 1.45 | 1.51 | 2.04 | 1.32 | 1.38 | 1.60 | 1.15 | 1.31 | 1.05 |
| MC25 | 1.48 | 1.60 | 2.00 | 1.40 | 1.17 | 1.55 | 1.47 | 1.32 | 1.05 |
| ωB97M-V | 1.14 | **1.01** | 1.07 | 1.12 | 1.12 | 1.78 | 0.92 | 1.32 | 0.96 |
| CF22D | 1.25 | 1.49 | 1.27 | **1.01** | 1.21 | 1.84 | 2.86 | 1.32 | 1.00 |
| B3LYP-D4 | 2.16 | 1.94 | 1.89 | 2.09 | 3.34 | 1.94 | 1.06 | 2.47 | 1.60 |
| PBE0-D4 | 1.97 | 2.45 | 1.62 | 1.83 | 2.85 | **1.36** | **0.72** | 1.93 | 1.52 |

**Figure 10 | Category-mean NERs of COF26, MC26 and representative functionals for selected GSCDB137 subsets.**